\documentclass[runningheads]{llncs}
\usepackage{llncsdoc}
\usepackage{cite}
\usepackage{hyperref}
\usepackage{fancyvrb}
\usepackage{wrapfig}
\usepackage{tikz}
\usepackage{url}
\usepackage{alltt}
\usetikzlibrary{arrows}
\usepackage{amsmath,amssymb}

\newcommand{\marimba}{\textsf{Marimba}}

\begin{document}
\title{Marimba: A Tool for Verifying Properties of Hidden Markov Models\thanks{\small The final version of this paper was accepted in the 13th International Symposium on Automated Technology for Verification and Analysis (ATVA 2015). The final publication is available at Springer via \url{http://dx.doi.org/10.1007/978-3-319-24953-7_14}}}

\titlerunning{Marimba: A Tool for Verifying Properties of Hidden Markov Models}

\author{No\'e Hern\'andez\inst{1} \and Kerstin Eder\inst{3,4} \and Evgeni Magid\inst{3,4} \and Jes\'us Savage\inst{2} \and \mbox{David A.\ Rosenblueth\inst{1}}}

\authorrunning{N. Hern\'andez, K. Eder, E. Magid, J. Savage, and D.\ A.\ Rosenblueth}

\institute{Instituto de Investigaciones en Matem\'aticas Aplicadas y en Sistemas 
\and Facultad de Ingenier\'ia \\Universidad Nacional Aut\'{o}noma de M\'{e}xico, D.F., M\'{e}xico \\
\email{no\_hernan@ciencias.unam.mx, drosenbl@unam.mx, savage@servidor.unam.mx}
\and Department of Computer Science, University of Bristol, Bristol, BS8 1UB, UK\\
\and Bristol Robotics Laboratory, Bristol, BS16 1QY, UK\\
\email{Kerstin.Eder@bristol.ac.uk, Evgeni.Magid@bristol.ac.uk}
}

\maketitle
\begin{abstract}
The formal verification of properties of Hidden Markov Models (HMMs) is highly de\-si\-ra\-ble for gaining confidence in the correctness of the model and the correspon\-ding system. A significant step towards HMM verification was the development by Zhang et al. of a family of logics for verifying HMMs, called POCTL*, and its model checking algorithm. As far as we know, the verification tool we present here is the first one based on Zhang et al.'s approach. As an example of its effective application, we verify properties of a handover task in the context of human-robot interaction. Our tool was implemented in \textsc{Haskell}, and the experimental evaluation was performed using the humanoid robot~\textsc{Bert2}.
\end{abstract}
\section{Introduction}
A Hidden Markov Model (HMM) is an extension of a Discrete Time Markov Chain (DTMC) where the states of the model are hidden but the observations are visible. Typically, an HMM is studied with respect to the three basic problems examined by Rabiner in \cite{Rabiner:89}. However, to the best of our knowledge, no practical model checker exists for HMMs despite their broad range of applications, e.g., speech recognition, DNA sequence analysis, text recognition and robot control. We describe in this paper a tool for verifying HMM properties written in the Probabilistic Observation Computational Tree Logic* (POCTL*~\cite{Zhang:05}), and use this tool for verifying properties of a robot-to-human handover interaction. 

POCTL* is a specification language for HMM properties. It is a probabilistic version of CTL* where a set of observations is attached to the {\em next} operator. Zhang et al.~\cite{Zhang:05} sketched two model checking algorithms for
POCTL*, an ``automaton based'' approach, and a ``direct'' approach.
We opted for the direct approach for its lower time complexity. Noticeably, this approach produces a DTMC $\mathcal{D}$ and a Linear Temporal Logic (LTL) formula $\phi$, so the \textsc{PRISM}\cite{PRISM} model checker could be used to verify this property. Such a model checker follows the automata based approach whose complexity is doubly exponential in $|\phi|$ and polynomial in $|\mathcal{D}|$, whereas we implemented the direct method by Courcoubetis et al. \cite{Courcoubetis:95} whose complexity is singly exponential in $|\phi|$ and polynomial in $|\mathcal{D}|$, which is also the final complexity of our tool. This direct method repeatedly constructs a DTMC and rewrites an LTL formula, such that one temporal operator is removed each time while preserving the probability of satisfaction.

We have named our model checker \marimba{}. A marimba is a xylophone-like musical instrument that is popular in south-east Mexico and Central America. \marimba\ \cite{Hernandez:14} was implemented in \textsc{Haskell} and compiled with GHCi. Our tool is available for download from \url{https://github.com/nohernan/Marimba}. 

\section{Tool architecture and implementation}
\textsc{Haskell} was chosen to code this first version of \marimba\ since it allows us to work in a high-level abstract layer, by providing useful mechanisms like lazy evaluation and a pure functional paradigm. Furthermore, \textsc{Haskell} manages recursion efficiently; this is a valuable aspect because recursive calls are made continuously throughout the execution. As a future work, we consider coding \marimba\ in a language like \textsc{Java} and make it a symbolic model checker.

\marimba\ features a command-line interface. Furthermore, instead of working with a command window, a more user friendly and preferable execution is accomplished through the {\em Emacs} text editor extended with the \texttt{Haskell-mode}.

\subsection{Marimba's input and modules}
The first input is a \verb@.poctl@ file with the six elements of an HMM $\mathcal{H}$, namely a finite set of states $S$, a state transition probability matrix $A$, a finite set of observations $\Theta$, an observation probability matrix $B$, a function $L$ that maps states to sets of atomic propositions from a set $AP_\mathcal{H}$, and an initial probability distribution $\pi$ over $S$. The second input is a POCTL* state formula $\Phi$ typed in the command window according to the syntactic rules:
\[\arraycolsep=2pt\def\arraystretch{1.2}
\begin{array}{lll}
\Phi & ::= & \mathsf{true} \;\;|\;\; \mathsf{false} \;\;|\;\; a \;\;|\;\; (\neg\Phi) \;\;|\;\; (\Phi\vee\Phi) \;\;|\;\; (\Phi\wedge\Phi) \;\;|\;\; (\mathcal{P}_{\bowtie\, p}(\phi)),\\
\phi & ::= & \Phi \;\;|\;\; (\neg\phi) \;\;|\;\; (\phi\vee\phi) \;\;|\;\; (\phi\wedge\phi) \;\;|\;\; (\mathbf{X}_\mathsf{o}\phi) \;\;|\;\; (\phi\,\mathcal{U}^{\leq n}\phi) \;\;|\;\; (\phi\,\mathcal{U}\phi),
\end{array}
\]
where $a\in AP_\mathcal{H}$, $\mathsf{o}\in\Theta$, $n\in\mathbb{N}$, $p\in [0,1]$, and $\bowtie\,\in\{\leq,<,\geq,>\}$. In addition, we define $\mathbf{X}_{\Omega}\phi$ as a shorthand for $\bigvee_{\mathsf{o}\in\Omega}\mathbf{X}_\mathsf{o}\phi$ provided $\Omega\subseteq\Theta$. We examine below the six \textsc{Haskell} modules that constitute \marimba.

\textit{ModelChecker.hs} performs the initial computations of the model checker for POCTL*. It recursively finds a most nested state subformula of $\Phi$, not being a propositional variable, and the states of  $\mathcal{H}$ that satisfy it. On the one hand, finding the states satisfying a propositional subformula is straightforward. On the other hand, we invoke the module \textit{DirectApproach.hs} to obtain the states satisfying a probabilistic state subformula. Next, this module extends the labels of such states with a new atomic proposition $a$. In $\Phi$, the state subformula being addressed is replaced by $a$. The base case occurs when we reach a propositional variable, so we return the states that have it in their label.

\textit{DirectApproach.hs} transforms the HMM $\mathcal{H}$ into a DTMC $\mathcal{D}$, and removes from the specification the observation set attached to the {\em next} operator $\mathbf{X}$ by generating a conjunction of the observation-free $\mathbf{X}$ with a new propositional variable. Thus, we obtain an LTL formula that is passed, together with $\mathcal{D}$, to the module \textit{Courcoubetis.hs}. The new propositional variables are drawn from the power set of observations. Remarkably, it is not necessary to compute such a power set since the label of a state in $\mathcal{D}$ is easily calculated.

\textit{Courcoubetis.hs} implements a modified version of the method by Courcoubetis et al. to find the probability that an LTL formula is satisfied in a DTMC. In this module, when dealing with the $\mathcal{U}$ and $\mathcal{U}^{\leq n}$	 operators, we apply ideas from \cite{Rutten:2004} for computing a partition of states of $\mathcal{D}$. Moreover, to handle the $\mathcal{U}$ operator we have to solve a linear equation system. To that end, we use the \textit{linearEqSolver} library \cite{linearEqSolver}, which in turn executes the \textit{Z3} theorem prover \cite{DeMoura:08}. 

\textit{Lexer.hs} and \textit{Parser.hs} are in charge of the syntactic analysis of the input. Finally, \textit{Main.hs} is loaded to start \marimba. This module manages the interaction with the user, and starts the computation by passing control to \textit{ModelChecker.hs}. 

In a typical execution, \marimba\ prompts the user to enter a \verb@.poctl@ file path. Next, our tool asks whether or not the user wants to take into account the initial distribution in the computation of the probability of satisfaction. This choice corresponds to opposite ideas presented in \cite{Courcoubetis:95} and \cite{Zhang:05}, i.e., the method by Courcoubetis et al. uses the
initial distribution to define their probability measure, contrary to that defined by Zhang et al. Afterwards, a POCTL* formula has to be entered. \marimba\ returns the list of states satisfying this formula, and asks the user whether there are more formulas to be verified on the same model.

The \verb@.poctl@ file is simply a text file where the elements of an HMM are defined, e.g., the set of states is defined by the reserved word \verb@States@, and if the model consists of five states, we write \verb@States=5@. Likewise, POCTL* formulas have a natural writing, for example, $\mathcal{P}_{<0.1}(\mathbf{X}_{\{o_1\}}a)$ is typed as \verb@P[<0.1](X_{1}a)@.

\section{Verification of a human-robot interaction}
\begin{wrapfigure}{r}{0.34\textwidth}
 \centering
\vspace{-6mm}
\begin{tikzpicture}[->,>=stealth',shorten >=0.6pt,auto,node distance=1.725cm,
  thick,main node/.style={circle,draw,font=\sffamily\scriptsize,align=center, inner sep=-6pt,minimum size=1mm}, bend angle=10]

  \node[main node]	(1) [text width=1.5cm, label=above: {\scriptsize State 1}, label={[label distance=0.4cm]90:{\scriptsize $L(1)$=\{\texttt{rnh}\}}}]			   {Robot not hold};
  \node[main node]	(2) [right of=1, text width=1.5cm, label=above: {\scriptsize State 2}, label={[label distance=0.4cm]90:{\scriptsize $L(2)$=\{\texttt{rpu}\}}}] {Robot pick up};
  \node[main node] 	(3) [below of=1, text width=1.5cm, label=below: {\scriptsize State 4}, label={[label distance=0.4cm]-90:{\scriptsize $L(4)$=\{\texttt{ug}\}}}] {User\\ grab};
  \node[main node] 	(4) [below of=2, text width=1.5cm, label=below: {\scriptsize State 3}, label={[label distance=0.4cm]-90:{\scriptsize $L(3)$=\{\texttt{rh}\}}}] {Robot\\ hold};

  \path[every node/.style={font=\sffamily\footnotesize}]
    (1)	edge [loop, style={min distance=5mm,in=185,out=145,looseness=5}] 	node {} (1)
	    	edge [bend left] 	node {} (2)
	    	
    (2) 	edge [loop, style={min distance=5mm,in=-5,out=35,looseness=5}] 	node {} (2)
       	edge	 [bend left]     node {} (4)
       	
	(3) 	edge [loop, style={min distance=5mm,in=185,out=145,looseness=5}] 	node {} (3)
       	edge	 [bend left]     node {} (1)
       	
	(4) 	edge [loop, style={min distance=5mm,in=-5,out=35,looseness=5}] 	node {} (4)
       	edge	 [bend left]     node {} (3);
\end{tikzpicture}
 \vspace{-8mm}
 \caption{The labelled states involved in the basic handover process.}
 \vspace{-7mm}
 \label{fig:basicHMM}
\end{wrapfigure}
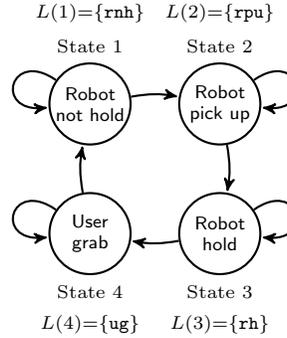

We applied \marimba{} to a real-world example, namely the verification of the robot-to-human handover task~\cite{Grigore:13} using the robot \textsc{Bert2}~\cite{Lenz:10} at the Bristol Robotics Laboratory (BRL). The robot's decision to release the object during the handover task is determined by an HMM \cite{Grigore:13}. Figure~\ref{fig:basicHMM} presents the state diagram of the HMM corresponding to the basic handover interaction, where the label $L(s)$ is defined for each state. 

Next, we initialise $A$, $B$ and $\pi$ of the HMM as follows. The process starts at state \textsf{Robot not hold}, so its initial distribution value $\pi_1$ is almost one, while the other states have initial distribution values close to zero. The initial matrix $A$ must encourage the transitions shown in Figure \ref{fig:basicHMM}. To initialise $B$, we consider
 as observations the ordered pairs whose first and second components are the index and middle finger metacarpophalangeal joint motor current values, respectively. By the Cartesian product of these va\-lues, we obtain 56,404 observations. Since these observations are merged with the states to generate the DTMC passed to {\em Courcoubetis.hs}, and the size of a formula could grow considerably by associating the {\em next} operator with up to 56,404 observations, \marimba{}'s execution is not practical under these circumstances. Vector quantisation\cite{Linde:80} was used to reduce the number of observations to just 13, which were taken to initialise matrix $B$. Thus, the initial ordered pairs are grouped into 13 regions of the plane representing the observations.

To make reliable estimates, we collected observations from 50 handover experiments on \textsc{Bert2}. These observations were used to train the initial HMM with the reestimation method found in the solution of Rabiner's Problem 3 \cite{Rabiner:89}.

\vspace{0.7em}
\noindent\textbf{Liveness properties.}
A liveness property requires that a {\em good thing} happens during the execution of a system. For example, we would like to know whether {\it the model generates the sequence of observations $\mathcal{O}=o_1, o_2, o_3, o_4$ where $o_1, o_2\in\{\mathsf{3}, \mathsf{4}, \mathsf{6}\}$ and $o_3, o_4\in\{\mathsf{3}, \mathsf{4}, \mathsf{11}\}$, with probability greater than 0.88}, that is,\linebreak
\(
\mathcal{P}_{>0.88}(\mathbf{X}_{\{3, 4, 6\}}(\mathbf{X}_{\{3, 4, 6\}}(\mathbf{X}_{\{3, 4, 11\}}(\mathbf{X}_{\{3, 4, 11\}}\mathsf{true}))))
\). Interestingly, this property is a generalisation of Rabiner's Problem 1 \cite{Rabiner:89}.
\marimba's execution for this property is found in Figure \ref{fig:Marimba}. The inputs are the trained HMM, defined in \texttt{ModelBert2.poctl}, and the previous formula. The output returned by \marimba{} is State $4$. Hence, the model starting at state \textsf{User grab} is likely to generate $\mathcal{O}$. 
\begin{figure}
\vspace{-3mm}
\begin{alltt}
Main> \textsl{main}
Enter the file name where the HMM is located.
\textsl{examples/ModelBert2.poctl}
Would you like to consider each state as if it were the initial 
state, i.e., as if it had initial distribution value equal to 1? y/n: \textsl{y}
Enter the POCTL* formula we are interested in.
\textsl{P[>0.88] (X_\{3,4,6\}(X_\{3,4,6\}(X_\{3,4,11\}(X_\{3,4,11\}T))))}
The states that satisfy it are:
(Probability of satisfaction of each state:[4.998198505964186e-10,
            4.08659792160621e-6,7.508994137303159e-3,0.8915357419467848])
[4]
Do you want to continue checking more specifications? y/n: \textsl{n}
\end{alltt}
\vspace{-4.6mm}
\caption{Verifying a property with \marimba{}.}
\vspace{-4.8mm}
\label{fig:Marimba}
\end{figure}

A second liveness property states that \textit{with probability at least 0.9, \textsc{Bert2} releases the object when the user grabs it.} The POCTL* formula for this property is
\(
\mathcal{P}_{\geq 0.9}(\mathtt{rh}\wedge(\mathtt{rh}\;\mathcal{U}\,(\mathtt{ug}\wedge\mathtt{ug}\;\mathcal{U}\,\mathtt{rnh})))
\).
\marimba{} outputs State 3, i.e., the specification is satisfied when the starting state is \textsf{Robot hold}. So, we expect \textsc{Bert2} to hold the object, and let it go when the user grabs it.

\vspace{0.7em}
\noindent\textbf{Safety properties.}
A safety property establishes that a {\em bad thing} does not occur during the execution of a system. For instance, \textit{with probability less than 0.05, \textsc{Bert2} abandons its serving position with the user not grabbing the object}, that is,
\(
\mathcal{P}_{<0.05}(\mathtt{rh}\wedge\mathbf{X}_{\Theta}(\mathtt{rnh}\vee\mathtt{rpu}))
\), where $\Theta$ is the set of observations.
Our model checker returns $\{1,2,3,4\}$ as the set of states satisfying this property. We conclude that it is unlikely that the model, being at state \textsf{Robot hold}, reaches a state other than \textsf{User grab}, that is, \textsf{Robot not hold} or \textsf{Robot pick up}.

The satisfaction of the previous three specifications provides us with confidence that \textsc{Bert2} reliably performs the handover interaction specified above.

On an Intel\textsuperscript{\textregistered} Core\texttrademark{} i3 1.70GHz computer with 4GB in memory, \marimba\ takes 28.55s to compute the states satisfying the first liveness formula. The time required for checking the other two properties studied here is around 0.06s.

Further examples are given in the {\em examples} folder and {\em user's manual} that come with \marimba's source code.

\section{Conclusions}
Since the automatic verification of properties of HMMs seems to be an unattended problem, we present here \marimba{}, a \textsc{Haskell} implementation of the model checking algorithm for POCTL*~\cite{Zhang:05}. This model checking algorithm was slightly modified to carry out its computations in a real program. \marimba's calculation is basically broken out in three stages that are coded in the modules {\em Mo\-delChecker.hs}, {\em DirectApproach.hs} and {\em Courcoubetis.hs}, such that the involved components, steps and transformations are well arranged throughout the implementation. Finally, we have successfully applied \marimba{} to verify relevant properties of a handover interaction from the robot \textsc{Bert2} to a human.

\vspace{0.7em}
\noindent\textbf{Acknowledgements.} We gratefully acknowledge support from grants PAPIIT IN113013 and Conacyt 221341, and especially thank the BRL staff for their assistance operating the robot \textsc{Bert2}. E. Magid and K. Eder have been supported, in full and in part, respectively, by the UK EPSRC grant EP/K006320/1 ROBOSAFE: ``Trustworthy Robotic Assistants".

\bibliographystyle{amsplain}
\bibliography{file}

\newpage
\section{\appendixname}
Technical details and formal definitions concerning HMMs and the POCTL* formalism are presented next.

\subsection{Hidden Markov Model}
An HMM has two layers, one on top of the other. The stochastic process between states on the underlying layer is hidden, and can be seen only through the stochastic process on the external layer that effectively produces a visible sequence of observations.
\begin{definition}
A labelled Hidden Markov Model {\em \cite{Rabiner:89}} consists of a tuple \linebreak$\mathcal{H}=(S,A,\Theta, B,L,\pi)$, where:
\begin{itemize}
\item $S=\{S_0,S_1,\ldots,S_{n-1}\}$ is a finite set of states;
\item $A$ is a state transition probability matrix, such that:
\[
A=\{a_{ij}\},\ a_{ij}\geq 0 \quad\; 0\leq i,j\leq n-1, \quad\qquad\sum\nolimits_{j=0}^{n-1} a_{ij} = 1 \quad\; 0\leq i\leq n-1;
\]
\item $\Theta=\{v_0,v_1,\ldots ,v_{m-1}\}$ is a set of $m$ observations;
\item $B$ is the observation probability matrix, $B=\{b_j(k)\}$ with 
\[
b_j(k)=P[v_k \,|\, S_j], \quad 0\leq j\leq n-1, \quad 0\leq k\leq m-1;\\
\]
\item $L:S\rightarrow 2^{AP_\mathcal{H}}$ maps states to sets of atomic propositions from a set $AP_\mathcal{H}$;
\item $\pi$ is an initial probability distribution over $S$, such that:
\[
\pi_i=P[q_0=S_i]\geq 0\qquad\, 0\leq i\leq n-1, \qquad\qquad \sum\nolimits_{i=0}^{n-1} \pi_{i}=1.
\]
\end{itemize}
\end{definition}

An execution of the system which is being modelled by an HMM is represented by a path.
\begin{definition}
A path {\em \cite{Zhang:05}} is a sequence $(s_0, o_0), (s_1, o_1), \ldots$, where $s_i\in S$, $o_i\in \Theta$, $a_{s_is_{i+1}}>0$ and $b_{s_i}(o_i)>0$, $\forall i\geq 0$. A path can be finite ($\omega^\mathsf{fin}$) or infinite ($\omega$).
\end{definition}

We denote the $(i+1)$st state of $\omega$ by $\omega_s(i)$, and the $(i+1)$st observation by $\omega_o(i)$. 
The suffix $(s_i,o_i), (s_{i+1},o_{i+1}), \ldots$ of $\omega$ is denoted by $\omega[i]$. We denote the sets of all finite and infinite paths in $\mathcal{H}$, starting with a pair whose state is $s$, by $\mathsf{Path}^{\mathsf{fin}, \mathcal{H}}_s$ and $\mathsf{Path}_s^\mathcal{H}$, respectively. 

To quantify the probability that an HMM behaves in a certain way, we define the measure $\mathrm{Pr}_s$ over the set $\mathsf{Path}_s^\mathcal{H}$. The {\em basic cylinder set} induced by the cylinder $\omega^{\mathsf{fin}}=(s_0,o_0), (s_1,o_1), \ldots, (s_k,o_k)$ is defined as
\(
C(\omega^{\mathsf{fin}})=\{\omega\in \mathsf{Path}_s^\mathcal{H}\;|\; \forall i\in\{0,\ldots,k\} \;(\omega_s(i)=s_i \wedge \omega_o(i)=o_i)\}.
\)
Let $\Sigma_s$ be the smallest $\sigma$-algebra on $\mathsf{Path}_s^\mathcal{H}$ which contains all basic cylinder sets $C(\omega^{\mathsf{fin}})$, where $\omega^{\mathsf{fin}}=(s,o_0), \ldots, (s_k,o_k)\in\mathsf{Path}^{\mathsf{fin}, \mathcal{H}}_s$. We define $\mathrm{Pr}_s$ on $\Sigma_s$ as,
\[
\mathrm{Pr}_s\Big(C\big((s,o_0), \ldots, (s_k,o_k)\big)\Big) = \pi_sb_s(o_0)\prod\nolimits_{i=1}^{k}a_{s_{i-1}s_i}b_{s_i}(o_i).
\]

Let $\Sigma$ be the smallest $\sigma$-algebra on $\mathsf{Path}^\mathcal{H}$ containing all basic cylinder sets, such that $\mathsf{Path}^\mathcal{H}$ is the set of paths in $\mathcal{H}$ with no constraint on the state of the initial pair. \mbox{In \cite{Hernandez:14}, the probability measure $\mathrm{Pr}_{\mathcal{H}}$ on $\Sigma$ is defined in terms of $\mathrm{Pr}_s$.}

We quantify the probability that an HMM behaves in a certain way by identifying the set of paths that satisfy a formula, and then using $\mathrm{Pr}_\mathcal{H}$ (or $\mathrm{Pr}_s$).

\subsection{POCTL*}
The Probabilistic Observation Computational Tree Logic* (POCTL*~\cite{Zhang:05}) has a {\em next} operator equipped with an observation constraint. 

\begin{definition}[Syntax]
Let $\mathcal{H}=(S,A,\Theta,B,L,\pi)$ be an HMM defined over the set of atomic propositions $AP_\mathcal{H}$. The syntax of POCTL* is defined as follows:
\[\arraycolsep=2pt\def\arraystretch{1.2}
\begin{array}{lll}
\Phi & ::= & \mathsf{true} \;\;|\;\; \mathsf{false} \;\;|\;\; a \;\;|\;\; (\neg\Phi) \;\;|\;\; (\Phi\vee\Phi) \;\;|\;\; (\Phi\wedge\Phi) \;\;|\;\; (\mathcal{P}_{\bowtie\, p}(\phi)),\\
\phi & ::= & \Phi \;\;|\;\; (\neg\phi) \;\;|\;\; (\phi\vee\phi) \;\;|\;\; (\phi\wedge\phi) \;\;|\;\; (\mathbf{X}_\mathsf{o}\phi) \;\;|\;\; (\phi\,\mathcal{U}^{\leq n}\phi) \;\;|\;\; (\phi\,\mathcal{U}\phi),
\end{array}
\]
where $a\in AP_\mathcal{H}$, $\mathsf{o}\in\Theta$, $n\in\mathbb{N}$, $p\in [0,1]$, and $\bowtie\,\in\{\leq,<,\geq,>\}$. We distinguish between state formulas $\Phi$ and path formulas $\phi$. 
\end{definition}
\begin{definition}[Semantics]
Let $\mathcal{H}=(S,A,\Theta,B,L,\pi)$ be an HMM. For any state $s\in S$, the satisfaction relation $\models$ is inductively defined as
\begin{multicols}{2}
\(\arraycolsep=2pt\def\arraystretch{1.2}
\begin{array}{lll}
s\models \mathsf{true}		&\multicolumn{2}{l}{\forall s\in S,}\\
s\not\models \mathsf{false}\;	&\multicolumn{2}{l}{\forall s\in S,}\\
s\models a 					& \mbox{iff}\; & a\in L(s),\\
s\models \neg\Phi 			& \mbox{iff} & s\not\models\Phi,
\end{array}\quad\quad
\begin{array}{lll}
s\models \Phi_1\vee\Phi_2		& \mbox{iff} & s\models\Phi_1 \vee s\models\Phi_2,\\
s\models \Phi_1\wedge\Phi_2	& \mbox{iff} & s\models\Phi_1 \wedge s\models\Phi_2,\\
s\models \mathcal{P}_{\bowtie\, p}(\phi)\; & \text{iff}\; & \mathrm{Pr}_s\{\omega\in\mathsf{Path}_s^\mathcal{H} \,|\, \omega\models\phi\} \bowtie p.
\end{array}
\)
\end{multicols}
\noindent For any path $\omega$, the satisfaction relation is defined as
\[
\arraycolsep=2pt\def\arraystretch{1.2}
\begin{array}{llllll}
\omega\models\Phi				& \mbox{iff} & \omega_s(0)\models\Phi,\quad\qquad\quad				& 
\omega\models\phi_1\vee\phi_2		& \mbox{iff} & \omega\models\phi_1 \vee \omega\models\phi_2, 	\\
\omega\models\neg\phi			& \mbox{iff} & \omega\not\models\phi,							&
\omega\models\phi_1\wedge\phi_2	& \mbox{iff} & \omega\models\phi_1 \wedge \omega\models\phi_2,\\
\multicolumn{6}{l}{
\hspace{-0.6mm}
\begin{array}{lll}
\omega\models\mathbf{X}_\mathsf{o}\phi& \mbox{iff} & \omega_o(0)=\mathsf{o} \wedge \omega[1]\models\phi, \\
\omega\models\phi_1\,\mathcal{U}^{\leq n}\phi_2\; & \mbox{iff}\; & \exists j \leq n.\;(\omega[j]\models\phi_2 \wedge \forall i<j.\;\omega[i]\models\phi_1), \\
\omega\models\phi_1\,\mathcal{U}\phi_2 & \mbox{iff} & \exists j \geq 0.\;(\omega[j]\models\phi_2 \wedge \forall i<j.\;\omega[i]\models\phi_1).
\end{array}
}
\end{array}
\]
\end{definition}

Let $\Omega\subseteq\Theta$, we write $\mathbf{X}_\Omega\phi$ as a shorthand for $\bigvee_{\mathsf{o}\in\Omega}\mathbf{X}_\mathsf{o}\phi$. Therefore, \linebreak$\omega\models\mathbf{X}_\Omega\phi$ iff $\omega_o(0)\in\Omega \wedge \omega[1]\models\phi$.

\section{Model checking algorithm}
Let \(\mathcal{H}=(S, A, \Theta, B, L, \pi)\) be an HMM, $s$ be a state in $S$, and $\Phi$ be a POCTL* state formula. Next, we explain a method to know whether $s\models\Phi$ holds or not.

\subsection{Stage One}
According to \cite{Zhang:05}, the model checking algorithm starts by taking a most deeply nested state subformula $\Psi$ of $\Phi$, such that $\Psi$ is not an atomic proposition. It is straightforward to find the states in $S$ that satisfy $\Psi$ when it is propositional. To obtain the states that satisfy $\Psi$ when it is of the form $\mathcal{P}_{\bowtie\, p}(\phi)$, stage two of the model checker is invoked. Once we determine the states satisfying $\Psi$, their label is extended by a new atomic proposition $a_\Psi$. Next, $\Psi$ is replaced by $a_\Psi$ in $\Phi$. The algorithm proceeds recursively, unless $\Phi$ itself is replaced by $a_\Phi$; in such case the algorithm returns states $s$, with $a_\Phi\in L(s)$. 

\subsection{Stage Two}
To identify the states that satisfy $\mathcal{P}_{\bowtie\, p}(\phi)$,  we follow the direct approach that transforms the original $\mathcal{H}$ into a DTMC $\mathcal{D}=(S^\mathcal{D}, A^\mathcal{D}, L^\mathcal{D}, \pi^\mathcal{D})$, where
\begin{multicols}{2}
\begin{itemize}
\item $S^\mathcal{D}= S\times\Theta$, 
\item $A^\mathcal{D}((s,o), (s',o'))=a_{ss'}\cdot b_{s'}(o')$, 
\item $L^\mathcal{D}(s,o)=L(s)\cup\{\Omega\subseteq\Theta\,|\,o\in\Omega\}$,
\item $\pi^\mathcal{D}_{(s,o)}=\pi_s\cdot b_s(o)$,
\end{itemize}
\end{multicols}
\noindent that is defined over the set of atomic propositions \mbox{$AP_\mathcal{D}=AP_\mathcal{H}\cup\{\Omega \,|\, \Omega\subseteq\Theta\}$}. The argument of $\mathcal{P}$, i.e., $\phi$, is modified to obtain $\phi'$ in a way that every occurrence of $\mathbf{X}_\Omega\varphi$ is replaced by $\Omega\wedge\mathbf{X}\varphi$. Notice that $\Omega$ is a new atomic proposition defined in $AP_\mathcal{D}$.

\subsection{Stage Three}
As stated in \cite{Courcoubetis:95, Hernandez:14}, stage three recursively constructs a new DTMC $\mathcal{D}'$ by applying the transformations $C_\mathbf{X}$, $C_\mathcal{U}$ and $C_{\mathcal{U}^{\leq n}}$, which are performed for each occurrence of $\mathbf{X}$, $\mathcal{U}$ and $\mathcal{U}^{\leq n}$, respectively. To show how the transformations work, we focus here on $C_\mathbf{X}$. It takes $\mathbf{X}\varphi$ as an innermost subexpression of $\phi'$. Then, it partitions the states of $\mathcal{D}$ into three disjoint subsets, $S^\mathcal{D}=S^\mathsf{YES}\cup S^\mathsf{NO}\cup S^\mathsf{?}$, where:
\begin{itemize}
\item $S^\mathsf{YES}$ consists of the states whose transitions are only into states satis\-fying~$\varphi$.
\item $S^\mathsf{NO}$ consists of the states whose transitions are only into states satisfying~$\neg\varphi$.
\item $S^\mathsf{?}$ consists of the states with transitions to both states satisfying $\varphi$ and states satisfying $\neg\varphi$.
\end{itemize}
Let $q_u$ denote the probability that $\mathbf{X}\varphi$ is satisfied starting from state $u\in S^\mathcal{D}$. We know that $q_u=1$ if $u\in S^\mathsf{YES}$, and $q_u=0$ if $u\in S^\mathsf{NO}$. Otherwise, $q_u=\sum_v A^\mathcal{D}(u,v)$, where the sum ranges over all successor states $v$ of $u$ satisfying formula $\varphi$. Let $\overline{q_u}=1-q_u$. Moreover, the new DTMC $\mathcal{D}'$ is defined over the set $AP_{\mathcal{D}'} = AP_\mathcal{D}\cup\{\xi\}$, where $\xi$ is a new atomic proposition representing $\mathbf{X}\varphi$.

\vspace{0.7em}
\noindent\textbf{States of $\boldsymbol{\mathcal{D}'}$.}
For each $u\in S^\mathsf{YES}$ there is a new state $(u,\xi)$ in $\mathcal{D}'$. For each $u\in S^\mathsf{NO}$ there is a new state $(u,\neg\xi)$. And for each $u\in S^\mathsf{?}$, there are two new states $(u,\xi)$ and $(u,\neg\xi)$. We define $L^{\mathcal{D}'}(u,\xi)=L^\mathcal{D}(u)\cup\{\xi\}$ and $L^{\mathcal{D}'}(u,\neg\xi)=L^{\mathcal{D}}(u)$.

\vspace{0.7em}
\noindent\textbf{Transitions of $\boldsymbol{\mathcal{D}'}$.}
The transition probability of $(u,\xi_1)\rightarrow(v,\xi_2)$, with $\xi_i\in\{\xi,\neg\xi\}$ and $i\in\{1,2\}$, is defined as being equal to the probability that $\mathcal{D}$, being at state $u$, transitions next to state $v$, and starting from state $v$ onward satisfies property $\xi_2$, conditioned on the event that in state $u$ it satisfies property $\xi_1$. 

\vspace{0.7em}
\noindent\textbf{Initial distribution of $\boldsymbol{\mathcal{D}'}$.}
If $u\in S^\mathsf{YES}\cup S^\mathsf{NO}$, then $\pi^{\mathcal{D}'}_{(u, \xi_1)}=\pi^{\mathcal{D}}_u$. If $u\in S^\mathsf{?}$, then there are two states in $\mathcal{D}'$ for $u$, namely $(u,\xi)$ and $(u,\neg\xi)$, with initial probabilities $\pi^{\mathcal{D}}_u\cdot q_u$ and $\pi^{\mathcal{D}}_u\cdot \overline{q_u}$, respectively. Furthermore, $\psi$ is obtained by replacing $\mathbf{X}\varphi$ by $\xi$ in $\phi'$.

If $\phi'$ originally has $k$ temporal operators, the algorithm applies $k$ times the appropriate transformations $C_\mathbf{X}$, $C_\mathcal{U}$ and $C_{\mathcal{U}^{\leq n}}$, to finally get the DTMC $\mathcal{D}^k$ and the propositional formula $\psi^k$. It is proved in \cite{Hernandez:14} that $s\models\mathcal{P}_{\bowtie p}(\phi)$ iff
\[
\displaystyle\sum_{o\in \Theta}\Bigg( \displaystyle\sum_{
	\begingroup
	\everymath{\scriptstyle}
	\small
	\substack{
	\xi_{i_1}\in\{\xi_1,\neg\xi_1\} \\
	\vdots \\
	\xi_{i_k}\in\{\xi_k,\neg\xi_k\}
	}
	\endgroup
	} \mathrm{Pr}_{\underbrace{\scriptstyle((\ldots((s,o),\xi_{i_1}),\ldots),\xi_{i_k})}_{\sigma_0}}\{\sigma\in\mathsf{Path}^{\mathcal{D}^k}_{\sigma_0} 	\,|\, \sigma\models\psi^k\}\Bigg) \bowtie p.
\]
Since $\psi^k$ is a propositional formula, $\mathrm{Pr}_{\sigma_0}\{\sigma\in\mathsf{Path}^{\mathcal{D}^k}_{\sigma_0}\,|\, \sigma\models\psi^k\}$ is $\pi^{\mathcal{D}^k}_{\sigma_0}$.

\end{document}